\documentclass[pra,nofootinbib]{revtex4}

\pdfoutput=1

\usepackage{amsmath,amssymb}
\usepackage{epsfig}
\usepackage{color}
\newcommand{\n}{\nonumber}
\newcommand{\be}{\begin{equation}}
\newcommand{\ee}{\end{equation}}
\newcommand{\bea}{\begin{eqnarray}}
\newcommand{\eea}{\end{eqnarray}}

\begin{document}

\title{A class of exactly solvable Convection-Diffusion-Reaction equations in similarity form with intrinsic supersymmetry}

\author{Choon-Lin Ho}

\affiliation
{Department of Physics, Tamkang University, Tamsui 25137, Taiwan}



\begin{abstract}

 In this work we would like to point out the possibility of generating a class of exactly solvable  convection-diffusion-reaction equation
 in similarity form with intrinsic supersymmetry, i.e., the solution  and the diffusion coefficient of the  equation are supersymmetrically related through their similarity scaling forms.

\end{abstract}



 \maketitle 



\section{Introduction}

Stochastic phenomena are rather ubiquitous in the world we live in. 
Most of the stochastic processes could be described by 
the Convection-Diffusion-Reaction (CDR)  equation. This  is an important type of second order differential equation which has found 
many important applications in physics, chemistry, astrophysics, engineering, and biology. 
 It is mainly employed to model stochastic phenomena that involve the change of concentration/population of one or more substances/species distributed in space  under the influence of three processes: 
 convection/drifting under the influence of external forces, diffusion which causes the substances/species  to spread in space, and  
 local reaction which modify the concentration/population \cite{M,GK1,GK2,HM,CP,V,dO,YW}.  
 In the absence of either the reaction force or the drift force, the CDR equation reduces to the well-known 
  Fokker-Planck equation (FPE)\cite{R,F} and the reaction-diffusion equation (RDE)  \cite{M}, respectively.
  
As with any equation in science, exact solutions of CDR equations are not easy to obtain in general.  
This is  reflected by the fact that many recent works on CDR equations are based on approximate and/or numerical methods.
Nevertheless, it is worthwhile to look for any method that helps find exact solutions of CDR equations.

In our previous work, we have considered constructing exactly solvable (ES) CDR equations (including FPE and RDE as special cases) using two symmetry methods, namely, the similarity method \cite{Ho1} and the supersymmetry method \cite{Ho2}. For details please refer to the references mentioned.  In this work we shall consider obtaining ES CDR systems by combining  the two symmetry methods.

In Ref. [12]  we have briefly described how 
to  generate similarity solutions of a class of CDR from the similarity solutions of another CDR equation through the supersymmetry (Darboux) transformations \cite{susy}.  There the supersymmetric (SUSY) connection is extrinsic, in that the transformation connects two different CDR equations. In this note we would like to point out the possibility of generating a class of exactly solvable CDR equation in similarity form with intrinsic supersymmetry, i.e., the solution  and the diffusion coefficient of the CDR equation  are supersymmetrically related through their similarity scaling forms.

\section{CDR in similarity form}

Let the concentration/population of a species at the position  $x$ at time $t$ be described by a function $P(x,t)$.
The one-dimensional CDR equation satisfied by $P(x,t)$ is given by{ \cite{M}
\be
\frac{\partial P(x,t)}{\partial t}= -\frac{\partial}{\partial x}\left(C(x,t)\,P(x,t)\right) 
+\frac{\partial^2}{\partial x^2}\left(D(x,t)\,P(x,t)\right) + R(P,x,t),
\label{CDR equation0}
\ee
with the convection coefficient  $C(x,t)$,  the diffusion coefficient $D(x,t)$, and  the
reaction term $R(P,x,t)$.    The domains we shall consider in this paper are the whole real line or  the half line.  

If a  CDR equation possess scaling symmetry then its functional form is unchanged under the scale transformation
\be
x=\epsilon^a \,\bar{x}\;\;\;,\;\;\; t=\epsilon^b \,\bar{t},~~~ \epsilon, a, b :{\rm \ real\ constants}.
\ee
The  functions $C, D, R$ and $P$ then have the following scaling forms in terms of the similarity variable $z$:
\bea
P(x,t)&=&t^\mu y(z), ~~C(x,t)=t^\gamma\tau(z),\n\\
~~ D(x,t)&=&t^\delta \sigma(z), ~~ R(P,x,t)=t^\rho \rho(z).
\eea
 where 
  \be
z\equiv\frac{x}{t^{\alpha}}, ~~\mbox{where}~
\alpha=\frac{a}{b}\;\;\;,\; b\neq 0\; ,
\label{z}
\ee  
and $y(z), \tau(z), \sigma(z)$ and $\rho(z)$ are functions of $z$.
Scaling symmetry requires that the exponents are linked by \cite{Ho1}
\be
\gamma=\alpha-1,~~ \delta=2\alpha-1,~~\rho=\mu-1.
\ee
Hence $\alpha$ and $\mu$ are the only two independent scaling exponents of the CDR  equation.
With these, the CDR equation is reduced  to an ordinary differential equation
\be
\sigma y''+(2\sigma' +\alpha\,z-\tau )\,y' -(\tau'+\mu-\sigma^{\prime\prime})\, y+\rho=0.
\label{ODE}
\ee
Here ``prime" represents derivative with respect to $z$.

As mentioned in the Introduction, in this work  we would like to consider  generating a class of exactly solvable CDR in similarity form with intrinsic supersymmetry, i.e., the solution $P(x,t)$ and the diffusion coefficient $D(x,t)$ are supersymmetrically related in that their scaling functions $y(z)$ and $\sigma(z)$ are supersymmetric pair.

\section{CDR with $\mu=-\alpha$ and $\tau=2\sigma^\prime + \alpha\,z$}

We consider a class of CDR equations with $\mu=-\alpha$ and $\tau=2\sigma^\prime + \alpha\,z$.  Eq.\,({\ref{ODE}) becomes
\be
-y^{\prime\prime} + \frac{\sigma^{\prime\prime}}{\sigma}\,y - \frac{\rho}{\sigma}=0.
\label{ODE3}
\ee
For $ \rho=0$ this equation becomes
\be
-y^{\prime\prime} + \frac{\sigma^{\prime\prime}}{\sigma}\,y=0.
\ee
It is immediate to see that it has solution  $y(z)=\sigma(z)$. Since in this case the CDR equation is the FPE, so one looks for normalizable $\sigma(z)$ as in this case $P(x,t)$ is a probability density function.

Our main concern here is with the special choice
$\rho=-\Phi\,y$. This gives
\be
-y^{\prime\prime} + \frac{\sigma^{\prime\prime}+\Phi}{\sigma}\,y=0.
\ee
This equation can be recast into
\bea
&&-y(z)^{\prime\prime} + (V(z)-E) y(z)=0,\label{y} \\
&&-\sigma(z)^{\prime\prime} +(V(z)-E) \sigma(x) -\Phi(z)=0\label{sigma},
\eea
with a function $V(x)$ and a constant $E$.

For $\Phi(z)=0$, which is just the case $\rho=0$ mentioned before, we get an exactly solvable CDR equation with any normalizable $\sigma(z)$, i.e., $y(z)=\sigma(z)$.

Below we consider two special non-trivial cases of $\Phi(z)$.

\section{$\Phi(z)=\sigma(z) \Delta E$}

We take $\Phi(z)$ to be proportional to $\sigma(z)$ with a real constant $\Delta E$.
This leads to
\be
-\sigma(x)^{\prime\prime} +\left[V(x)-(E+\Delta\,E)\right] \sigma(x)=0.
\label{s1}
\ee
Now both Eq.(\ref{y}) and (\ref{s1}) are in the Schr\"odinger  form, with the same potential $V(z)$ and energies $E$ and $E+\Delta\,E$, respectively.

Take any exactly solvable Schr\"odinger equation 
\be
Hu_n=E_n u_n, ~~n=0,1,2,\ldots
\ee
Suppose we take $y(z)$ and $\sigma(z)$ to be eigenfunctions corresponding to two eigenvalues, say
$y(z)=u_n(z)$ and  $\sigma(z)=u_{m}(z)$ with $E=E_n$ and  $E+\Delta\,E=E_m$, respectively.
Then an ES CDR system is obtained, with its various functions given by
\bea
D(x,t)&=&t^{2\alpha-1}\,u_{m}(z),\n\\
C(x,t)&=&t^{\alpha-1}\left(2u_{m}(z)+\alpha z\right),\n\\
R(x,t)&=&-t^{-(\alpha-1)}\left(E_m- E_n\right) u_{m}(z)\,u_{n}(z),\n\\
P(x,t)&=& t^{-\alpha} u_{n}(z).
\eea

We note here that we can obtain another ES system by interchanging $m$ and $n$.

In the case $E_m=E_n$ (i.e., $\Delta\,E=0$), $y(z)=\sigma(z)$ as mentioned in the last section.


\section{$\Phi(z)=-\sigma(z) \Delta V(z)$}

Next we change $\Delta E$ to a function $\Delta V(z)$.
With this choice Eq.(\ref{sigma}) becomes
\be
-\sigma(z)^{\prime\prime} +(V(z)+\Delta\,V(z)-E) \sigma(z)=0.
\label{s1}
\ee
Both Eq.(\ref{y}) and (\ref{s1}) are in the Schr\"odinger  form.

So one could obtain an exactly solvable CDR by chossing $y(z)$ and $\sigma(z)$ to be solutions of the respective Schr\"odinger equations with the same energy $E$ but with different potentials $V(z)$ and $V(z)+\Delta(z)$, respectively.  

An easy way to accomplish this is to consider the two potentials to be the SUSY pair.  

The main ideas of SUSY QM relevant to our purpose here are summarized below (for details please see [13, 14]).

\subsection{Supersymmetry}

Consider the Schr\"odinger equation
\be
-\phi^{(0)}(x)^{\prime\prime} +( V_0(x) - E)\phi^{(0)}(x)=0.
\label{SE}
\ee
 Suppose $\phi^{(0)}(x)$ and $\phi^{(0)}_k(x)$ are solutions of (\ref{SE}) corresponding to eigenvalues $E$ and $E_k$ (for some index $k$), respectively.  Then the Darboux theorem states that 
 the set of functions defined by the following transformations,
 \bea
 V_1&=&V_0-2(\ln\phi^{(0)}_k)^{\prime\prime},\n\\
 \phi^{(1)}&=&\left(\partial_x  -(\ln\phi^{(0)}_k)^\prime \right)\phi^{(0)},
 \eea
 also satisfy the same form of Schr\"odinger equation with the same eigenvalue $E$,
 \be
-\phi^{(1)}(x)^{\prime\prime} +( V_1(x) - E) \phi^{(1)}(x)=0.
\ee 
It is customary to take $\phi^{(0)}_k$ to be the ground state $\phi^{(0)}_0$ of $V_0$.  The two potentials $V_0$ and $V_1$ are commonly called SUSY  pair in physics literature.  $V_1$ has  the same spectrum as that of $V_0$, except the ground state energy $E_0$ corresponding to $\phi_0^{(0)}$.

Now it is clear how to construct an exactly solvable CDR equation with intrinsic SUSY.  One just chose a SUSY pair of potentials $V_0(z)$ and $V_1(z)$ (now in variable $z$) and assign one set to the equation of $y(z)$ and the other to that of $\sigma(z)$.  Such examples are easily constructed, and will  not be dwelled on here.  Instead, we shall turn to SUSY potentials with a nice property -- shape invariance, a property possessed by most exactly solvable one-dimensional quantum systems. 

\subsection{Shape-invariant potentials}

Let's begin with a potential $V_0(x; \mathbf{a}_0)$  with eigenvalues $E_n^{(0)}(\mathbf{a}_0)$ and eigenfunctions $u_n^{(0)}(x, \mathbf{a}_0)$ ($n=0,1,2,\ldots$), where $\mathbf{a}_0$ is a set of parameters characterizing the system.  Shape invariance means that $V_0(x; \mathbf{a}_0)$ and
its SUSY partner potential $V_1(x; \mathbf{a}_0)$ are connected  by the realtion
\be
V_1(x; \mathbf{a}_0)=V_0(x; \mathbf{a}_1) + R(\mathbf{a}_0).
\ee
Here $\mathbf{a}_1$ is a function of $\mathbf{a}_0$ and $R(\mathbf{a}_0)$ is an $x$-independent function of $\mathbf{a}_0$.
This means $V_1(x; \mathbf{a}_0)$ is just  $V_0(x; \mathbf{a}_0)$ shifted by an constant $R(\mathbf{a}_0)$ and with the parameter $\mathbf{a}_0$ replaced by $\mathbf{a}_1$.
Thus the functional form of $V_1(x; \mathbf{a}_0)$ is similar to that of $V_0(x; \mathbf{a}_0)$.
Simple algebra shows that  the spectra of the two potentials are identical except the ground state of $V_0$, and the eigenfunctions of $V_1$ have the similar form as the corresponding ones of $V_0$ with $\mathbf{a}_1$ replacing $\mathbf{a}_0$, i.e.,
\be
E_n^{(1)}=E_{n+1}^{(0)}, ~~ u_n^{(1)}(x; \mathbf{a}_0)= u_n^{(0))}(x; \mathbf{a}_1).
\ee

It is worth noting that for  all the well-known SUSY one-dimensional solvable potentials,  $\mathbf{a}_1$ differs from $\mathbf{a}_0$ only by constant shifts, i.e., $\mathbf{a}_1=\mathbf{a}_0+\boldsymbol{\delta}$. 

The presence of shape invariance permits one to apply the SUSY transformation successively.  The $s$-step ($s=0,1,2,\ldots$)  SUSY partner potential, eigenvalues, and eigenfunctions are given by
\bea
V_s(x, \mathbf{a_0})&=&V_0(x, \mathbf{a}_s) + \sum_{k=0}^{s-1} R(\mathbf{a}_k)
\label{susy}
\\
E_n^{(s)}&=&E_{n+s}^{(0)},~~~~ u_n^{(s)}(x; \mathbf{a}_0)= u_n^{(0))}(x; \mathbf{a}_s).\n
\eea
It is understood that the summation term in $V_s(a; \mathbf{a}_0)$ is absent for $s=0$.

Returning to CDR equation, we can now construct exactly solvable CDR equations by considering Eq.(\ref{y}) and (\ref{s1}) to be SUSY pair, i.e., by choosing $V(z)$ and $V(z)+\Delta\,V(z)$ to the the $s$ and $s^\prime$ members in the series of shape-invariant potentials,
\be
V(z) =V_s(z),~~ V(z)+\Delta\,V(z)=V_{s'}(z),
\ee
and $y(z)$ and $\sigma(z)$ the corresponding eigenfunctions 
\be
y(z)= A u_n^{(s)}(z), ~~ \sigma(z)=B u_{n'}^{(s')}(z),  
\ee
with the same energy, $E_n^{(s)}=E_{n'}^{(s')}$, or equivalently,  $E_{n+s}^{(0)} = E_{n'+s'}^{(0)}$. The last equation implies the constraint between the indices $n^\prime + s^\prime =n+s$.
The real constants $A, B$ are arbitrary, owing to the linearity of (\ref{y}) and (\ref{s1}).

The various functions of the CDR equation are
\bea
D(x,t)&=&B t^{2\alpha-1}\,u_{n'}^{(s')}(z),\n\\
C(x,t)&=&t^{\alpha-1}\left(2 B u_{n'}^{(s')}(z)+\alpha z\right),\label{fn}\\
R(x,t)&=&AB t^{-(\alpha-1)}\left[V_{s'}(z)-V_s(z)\right] u_{n'}^{(s')}(z)u_{n}^{(s)}(z),\n\\
P(x,t)&=& A t^{-\alpha} u_{n}^{(s)}(z).
\eea

A corresponding ES system is obtained by interchanging $y(z)$ and $\sigma(z)$, or
$(n,s, A) \Leftrightarrow (n^\prime, s^\prime, B)$.

We demonstrate this construction by the example of the radial oscillator.

\subsection{Example: The radial oscillator}

The potential for the $s=0$ member of the series of SUSY radial oscillator potentials is
\be
V_0(x; \mathbf{a}_0)=\frac14 \omega^2 x^2 + \frac{\ell+1}{x^2}-\omega\left(\ell+\frac32\right),
\ee
where $x\in [0,\infty)$ and $\omega,  \ell>0$. 
Eigenvalues are given by $E_n^{(0)}=2n\omega, ~~n=0,1,2,\ldots$.
The wavefunctions are
\be
u^{(0)}_n(x)\equiv =N_{n\ell} y^{\frac{\ell+1}{2}} e^{-\frac{y}{2}} L_n^{\ell+\frac12}(y), ~~ y\equiv \frac12 \omega x^2,
\ee
with the normalization constant
\be
N_{n\ell}= (2\omega)^\frac14 \left[\left( \begin{array}{c}
 n+\ell +\frac12\\
n
 \end{array} \right) \Gamma(\ell+\frac32)
 \right]^{-1},
\ee
where the two factors in the square-bracket are the binomial coefficient and the Gamma function, respectively. 

 In this case  $\mathbf{a}_0=(\omega, \ell)$ and $\boldsymbol{\delta}=(0, 1)$, i.e., $\mathbf{a}_s=(\omega, \ell+s)$ \cite{susy}.  The function $ R(\mathbf{a}_s)=2\omega$ is independent of $s$.  The potentials, energies, and wavefunctions of the $s$-member of the SUSY chains are given by (\ref{susy}).
 
In Fig.\,1 we plot the graphs of the $P(x,t), D(x,t), C(x,t)$ and $R(x,t)$ for $\alpha=\omega=\ell=1$, and two sets of parameters $(n, s, A)=(3, 1, 1)$ and 
$(n^\prime, s^\prime, B)=(1, 3, 3)$.  Fig.\,2 presents the corresponding graphs with the last two sets of parameters interchanged.  

It is clear that the graphs of $P(x,t)$ in Fig.\,1 (Fig.\,2)  and $D(x,t)$ in Fig.\,2 (Fig.\,1) look quite similar except their scales, as the wavefunctions are the same except being scaled by different time factors and the cponstants $A, B$.  The function $R(x,t)$ in the two figures differ by a sign as a result of the interchange of the two potentials in (\ref{fn}).

In summary,  we have shown how to generate a class of exactly solvable CDR equation in similarity form with intrinsic supersymmetry.

\section*{Acknowledgments}
The work is supported in part by the Ministry of Science and Technology (MOST)
of the Republic of China under Grants   NSTC 112-2112-M-032-007 and NSTC 113-2112-M-032-010.




\begin{figure}[ht] \centering
\includegraphics*[width=5cm,height=6cm]{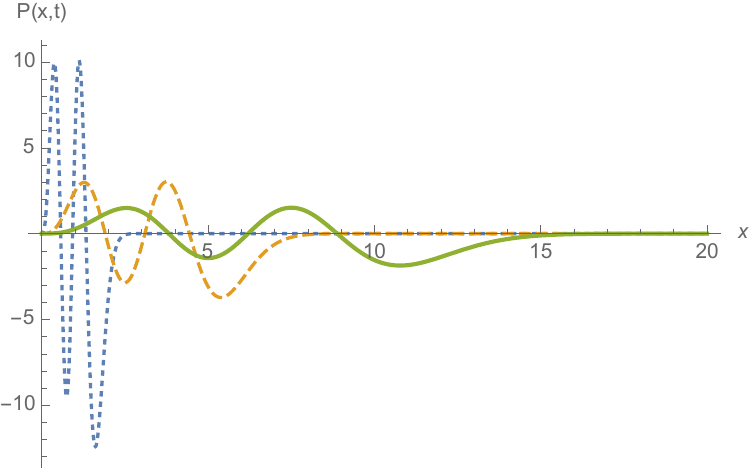}\hspace{1cm}
\includegraphics*[width=5cm,height=6cm]{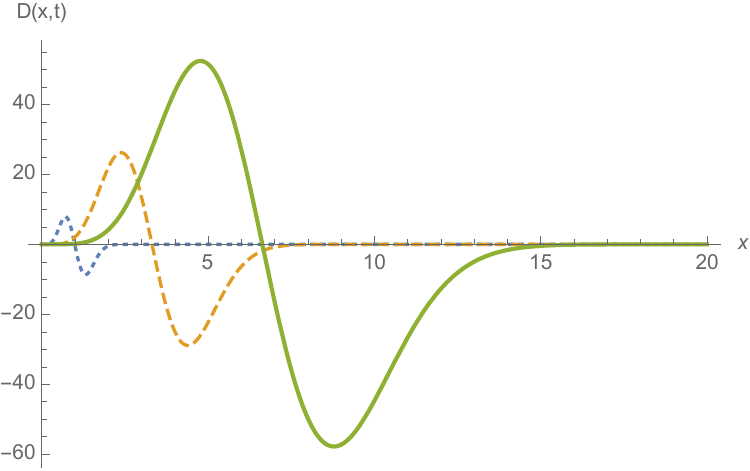}\\
\includegraphics*[width=5cm,height=6cm]{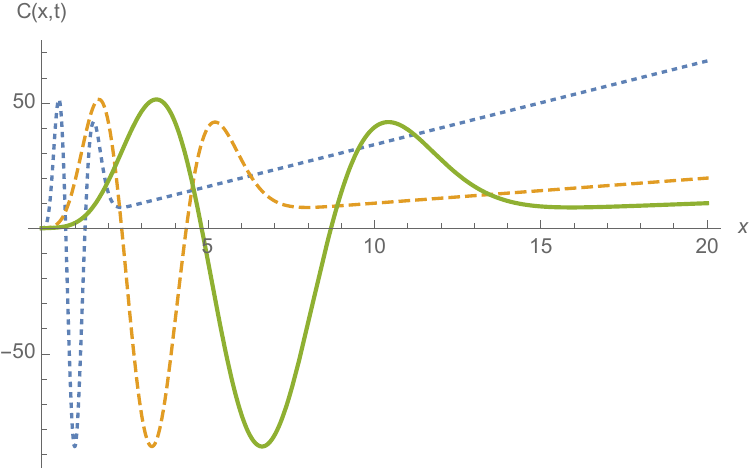}\hspace{1cm}
\includegraphics*[width=5cm,height=6cm]{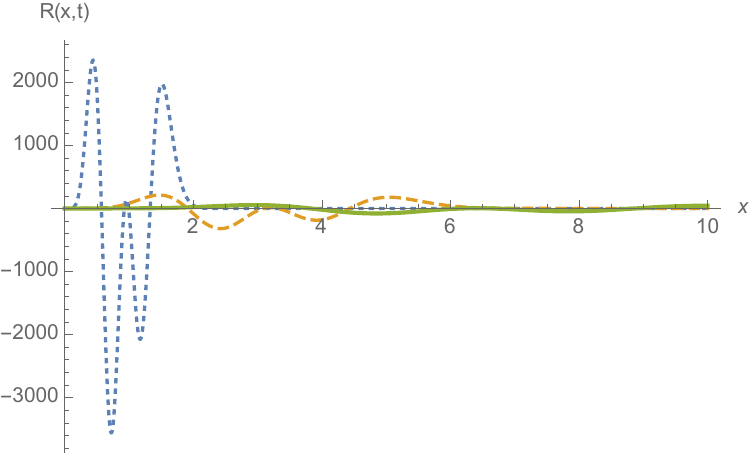}
\caption{Plot of $P(x,t), D(x,t),C(x,t), R(x,t)$ and $R(x,t)$ for the CDR system based on the radial oscillator discussed in Set.\,5.3. for time $t = 0.3$ (dotted), $t = 1.0$ (dashed), $t = 2.0$ (solid).The scaling exponent $\alpha=1$, and $A=1, B=3$. The parameters for the oscillator are $\omega=\ell=1$.  The members $s=1$ and $s^\prime =3$ with eigenstates $n=3$  and $n^\prime=1$, respectively,  are chosen to construct the CDR system.}
\label{Fig1}
\end{figure}

\begin{figure}[ht] \centering
\includegraphics*[width=5cm,height=6cm]{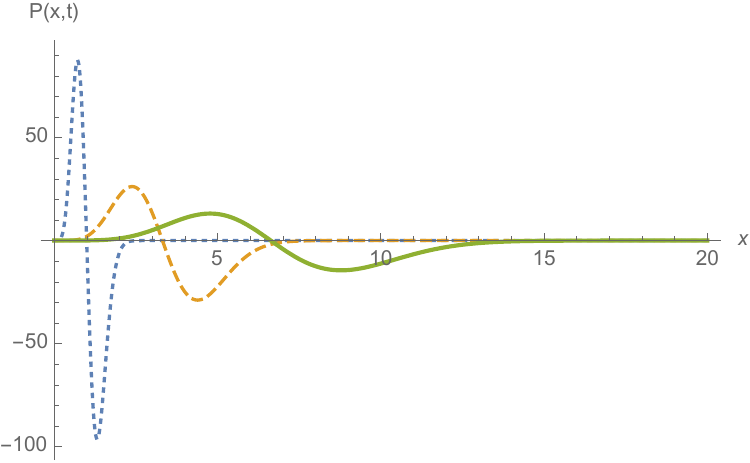}\hspace{1cm}
\includegraphics*[width=5cm,height=6cm]{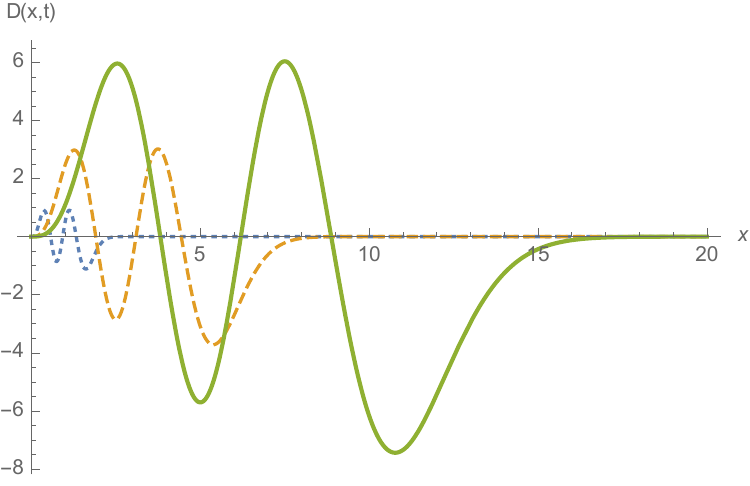}\\
\includegraphics*[width=5cm,height=6cm]{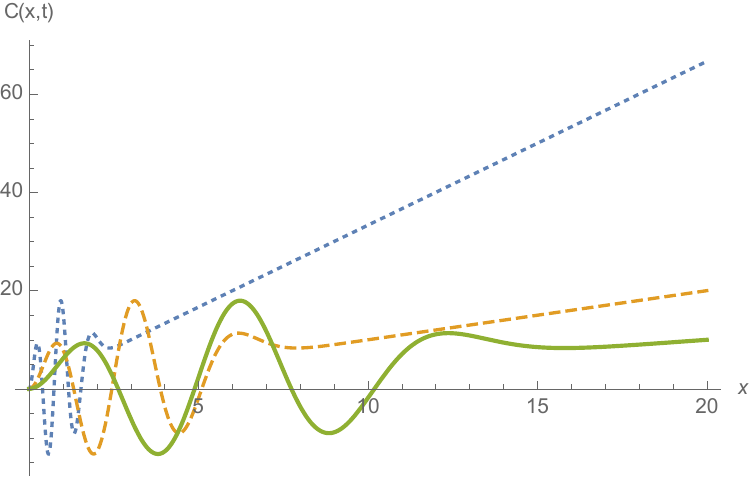}\hspace{1cm}
\includegraphics*[width=5cm,height=6cm]{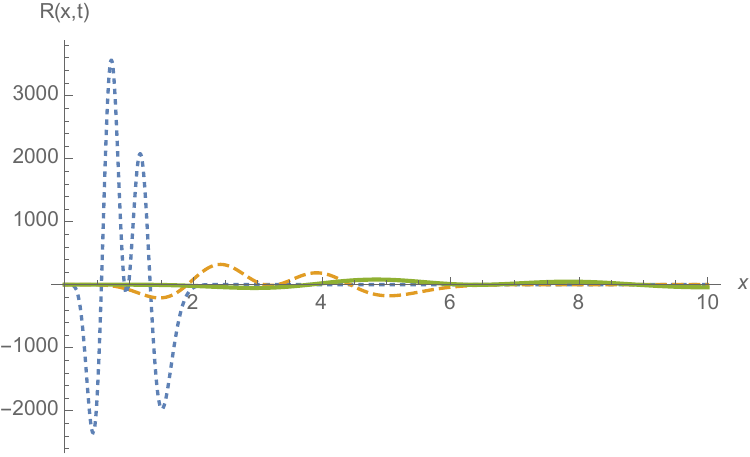}
\caption{Same as Fig.\,1 with the two sets of parameters $(n,s,A)$ and $(n^\prime, s^\prime, B)$ interchanged.}
\label{Fig2}
\end{figure}


\begin{thebibliography}{99}

\bibitem{M}
J.D. Murray, Mathematical Biology, 2nd Ed., Springer-Verlag, Berlin, 1993.

\bibitem{GK1}
B. H. Gilding and R. Kersner, Travelling Waves in Nonlinear Diffusion Convection Reaction, Birkh\"auser, Springer, 2004.

\bibitem{GK2}
B. H. Gilding and R. Kersner, 
The characterization of reaction-convection-diffusion processes by travelling waves, 
Journal of differential equations 124, 27 (1996).

\bibitem{HM}
T. Harko and M. K. Mak, 
 Exact travelling wave solutions of non-linear reaction-convection-diffusion equations: an Abel equation based approach,
 J. Math. Phys. 56, 111501 (2015).
   
\bibitem{CP}
R. Cherniha and O. Pliukhin,
New conditional symmetries and exact solutions of nonlinear reaction-diffusion-convection equations. I, II, and III.
arXiv:math-ph/0612078, arXiv:0706.0814, arXiv:0902.2290.

\bibitem{V}
E. Vidal-Henriquez, V. Zykov, E. Bodenschatz, and A. Gholami,
Convective Instability and Boundary Driven Oscillations in a Reaction-Diffusion-Advection Model,
Chaos 27, 103110 (2017).

\bibitem{dO}
L. M. de Oliveira Vilaca, B. Gomez-Vargas, S. Kumar, R. Ruiz-Baier, and N. Verma,
Stability analysis for a new model of multi-species convection-diffusion-reaction in poroelastic tissue,
Applied Mathematical Modeling, 84, 425 (2020).


\bibitem{YW}
K. Yamazaki and X. Wang,
Global stability and uniform persistence of the reaction-convection-diffusion cholera epidemic model,
Math. Biosci. Eng. 14, 559 (2017).


\bibitem{R}
H. Risken, The Fokker-Planck Equation, 2nd. ed., Springer-Verlag, Berlin, 1996.

\bibitem{F}
Sau Fa Kwok,
Langevin and Fokker-Panck Equations and Their Generalizations,
World Scientific, Singapore, 2018.


\bibitem{Ho1}
C.-L. Ho and C.-M. Yang, Convection-Diffusion-Reaction equation with similarity solutions, 
 Chin. J. Phys. 59, 117 (2019).
 
 \bibitem{Ho2}
 C.-L. Ho, 
Supersymmetry and convection-diffusion-reaction equations ,
 Int. J. mod. Phys.  B 38, 2450068 (2024).
 

\bibitem{susy} 
F. Cooper, A. Khare, and U. Sukhatme, Supersymmetry and quantum mechanics, Phys. Rep.  251, 267 (1995).
 
 \bibitem{Ho3}
 C.-L. Ho, Time-dependent Darboux transformation and supersymmetric hierarchy of Fokker-Planck equations,
Chin. J. Phys. 77, 1903 (2022), Appendix.

 
 \end{thebibliography}
\end{document}